# CDST: A Toolkit for Testing Cockpit Display Systems of Avionics

Hassan Sartaj*†, Muhammad Zohaib Iqbal*† and Muhammad Uzair Khan*†
*Quest Lab, Department of Computer Science, National University of Computer and Emerging Sciences, Islamabad
† UAV Dependability Lab, National Center of Robotics and Automation (NCRA)
Email: hassan.sartaj@questlab.pk, {zohaib.iqbal,uzair.khan}@nu.edu.pk

*Abstract*—Avionics are highly critical systems that require extensive testing governed by international safety standards. Cockpit Display Systems (CDS) are an essential component of modern aircraft cockpits and display information from the user application using various widgets. A significant step in the testing of avionics is to evaluate whether these CDS are displaying the correct information. A common industrial practice is to manually test the information on these CDS by taking the aircraft into different scenarios during the simulation. Given the large number of scenarios to test, manual testing of such behavior is a laborious activity. In this paper, we present a CDST toolkit that automates the testing of Cockpit Display Systems (CDS). We discuss the workflow and architecture of the tool and also demonstrates the tool on an industrial case study. The results show that the tool is able to generate, execute, and evaluate the test cases and identify 3 bugs in the case study.

## I. INTRODUCTION

Avionics software systems need to meet the quality requirements set by various international safety standards [1]. To meet the safety requirements of the standards, the testing and verification activities of avionics software require an extensive amount of effort and cost [2]. A significant enhancement to the modern-day aircraft is the introduction of a glass cockpit that comprises of a Cockpit Display Systems (CDS). CDS have replaced the dials and gauges in the aircrafts [3] and are considered as a significant component of the modern cockpits.

CDS display information that is vital for the safe operation of an aircraft. This includes information received from a number of user applications, the flight management system, flight control unit and the warnings generated by different hardware components. It is important to test that the information displayed on the CDS is correct. Such testing activity is largely done on simulators that simulate the various scenarios of aircraft operation. A common practice by the CDS testers is to test the information displayed on CDS by manually executing different aircraft scenarios and manually verifying that correct information is displayed according to these scenarios [4]. The scenarios are typically executed with the help of simulators. The evaluation of testing activities is done manually by the pilots or domain experts. This step has to be performed repeatedly whenever the required information to be displayed is changed, for example, due to an upgraded sensor being used. Testing in this way (manual execution and manual verification of results) is a very time consuming, error-prone, and laborious task. Moreover, the process of manual testing is not repeatable.

In our previous work [5], we proposed a model-based testing approach to automate the testing of CDS. In this paper, we focus on the automated testing tool that is used for testing the CDS. We present the CDST toolkit that consists of seven modules to assist the avionics engineers in the process of testing CDS, (i) *CDS Model Generator*, (ii) *Comparator*, (iii) *Reporting Module*, (iv) *CDS Constraint Specifier*, (v) *Test Path & Script Generator*, (vi) *Test Execution Module*, and (vii) *Cockpit Display Recorder*. The toolkit can be download from Github[1]. The toolkit takes as input the VAPS XT CDS interfaces, the behavior of the aircraft that is significant for testing modeled as a UML state machine, and OCL constraints specifying the ranges of possible values for various CDS elements during flight operations. The toolkit generates the various flight paths for testing, executes the test scripts on the simulator, performs test evaluation, and report results. We demonstrate the toolkit on an industrial case study. The results show that the toolkit is able to generate, execute, and evaluate the test cases and identify 3 bugs in the case study.

The remaining paper is organized as follows. Section II describes the CDS testing strategy that is used to develop CDST. Section III provides a detailed discussion on the CDST toolkit. Section IV presents the demonstration of the CDST toolkit. Section V discusses the limitations of the toolkit. Section VI describes related work. Finally, Section VII concludes the paper.

## II. CDS TESTING APPROACH OVERVIEW

This section gives an overview of the Cockpit Display Systems (CDS) testing approach presented in our previous work [5]. The approach provides the basis for the development of the CDST toolkit. As shown in Fig. 1, the approach requires three artifacts as input. The first artifact is the XML file of CDS modeled in a graphical modeling tool such as VAPS XT [6] or SCADE [7]. The second artifact is the behavior specification (state machine) of the possible states of an aircraft during the flight that has an impact on the information being displayed on CDS. The third input is the constraints written in Object Constraint Language (OCL) that are specified on different CDS elements according to aircraft flight states. To assist avionics test engineers in writing OCL constraints, we develop *CDS Constraint Specifier* tool as a part of the CDST toolkit.

---
[1]https://github.com/hassansartaj/cdst-toolkit

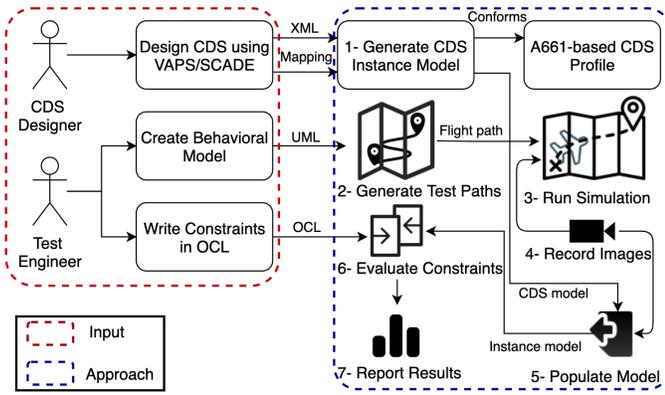

Fig. 1. An overview of the CDS testing approach

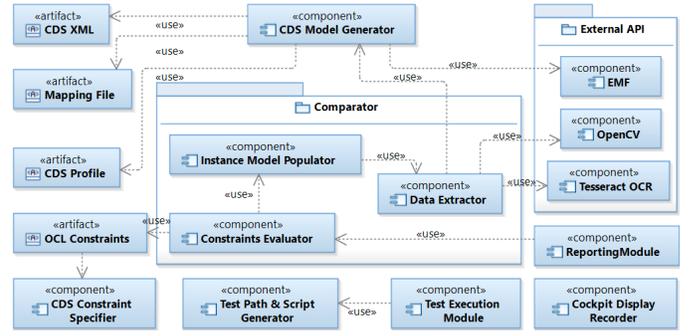

Fig. 2. Architecture diagram of CDST toolkit

The XML file representing CDS under test along with the CDS profile[2] and the mapping between them is used to generate the CDS model using *CDS Model Generator* module. The aircraft behavioral model is used to generate flight test paths. To automate the generation of flight test paths, we develop *Test Path & Script Generator* tool. According to each flight test path, the JSBSim simulator specific scripts are also generated by the tool. During the simulation, the data displayed on CDS is recorded using a camera or by taking screenshots after a specified time interval. For this purpose, the CDST toolkit contains the *Cockpit Display Recorder* tool. The images are stored according to the aircraft states. The generated CDS model is used to locate the CDS elements and to extract the data from images using the *Data Extractor* module. Using the CDS model and the data extracted from images, the instance models are populated using the *Instance Model Populator* module. Lastly, OCL constraints are evaluated on CDS instance models using the *Constraints Evaluator* module. The OCL constraints act as oracle during testing and provide the expected values for the various widgets of the CDS. If an instance model fails to satisfy an OCL constraint, the evaluator returns *false* and a potential bug is detected. The results are compiled by using the *Reporting Module*.

## III. CDST Toolkit

The CDST toolkit consists of seven modules to assist the avionics engineers in the process of testing CDS, (i) *CDS Model Generator*, (ii) *Comparator*, (iii) *Reporting Module*, (iv) *CDS Constraint Specifier*, (v) *Test Path & Script Generator*, (vi) *Test Execution Module*, and (vii) *Cockpit Display Recorder*. Fig. 2 shows the architecture diagram of the CDST toolkit. In the following, we discuss each module individually.

[2]https://github.com/hassansartaj/models19

### A. CDS Model Generator

The module *CDS Model Generator* generates the CDS model using the XML file of the CDS under test, CDS profile, and the mapping file. The CDS XML file is parsed using Java XML parser and the CDS profile (in UML format) is loaded using the Eclipse Modeling Framework (EMF). The mapping file contains the CDS element mentioned in the XML file as a source and the corresponding CDS profile element as a target. The purpose of the mapping file is to resolve the naming conflict between CDS elements designed using a tool (VAPS XT [6] or SCADE [7]) and CDS profile elements. Fig. 3 shows an example of the CDS model generation for the *AltitudeTape* of CDS. On the left-hand side, an excerpt from CDS XML is shown that represents the *AltitudeTape* of a primary flight display (PFD). In the center, an excerpt from the CDS profile is shown. On the right-hand side, the generated CDS model for the *AltitudeTape* part of PFD is shown. First, the root object name in CDS XML is matched with the CDS profile element. In this case, the object *AltitudeTape* is mapped to the profile element *Altimeter*. After a match is found, the nested properties in XML are mapped to the widget properties of the profile. As shown in Fig. 3, the property *IsVisible* is mapped to the same property in the profile widget. Similarly, the XY properties related to position and size are mapped. The mapping between CDS XML and profile elements is performed using the *XML2ProfileMapper*. At the end of mapping, the CDS model (an instance of the profile) is generated using EMF (as shown on the right-hand side of Fig. 3). The CDS model contains the information required to identify the *AltitudeTape* in CDS screen and the target property (i.e., *altitude*) whose value is used during test evaluation.

### B. Comparator

The function of the *Comparator* module is to prepare the test evaluation environment, perform test evaluation, and report results. To prepare the test evaluation environment, the sub-modules *Data Extractor*, *Instance Model Populator*, *Constraints Evaluator*, and *Reporting Module* play a key role.

*1) Data Extractor:* The first important sub-module is *Data Extractor*. Its main purpose is to extract the data from images recorded during the simulation and provide the extracted data to the *Instance Model Populator* module. To extract data from images, the CDS model generated in the previous step is used. The CDS model provides complete information about each widget on CDS. For example, the position (x and y-axis) of the widget, size, and color information. This information is used to identify various widgets in the image. Using the information obtained from the CDS model, the subpart of the image containing the target widget is located. To extract the subpart of the image, region-based segmentation is applied using OpenCV 3.4.1 [8] Java API. The image subpart is processed to remove noise and is fed into optical character recognition (OCR) software to extract the text. To perform optical character recognition (OCR) in the image, we use Tesseract OCR [9].

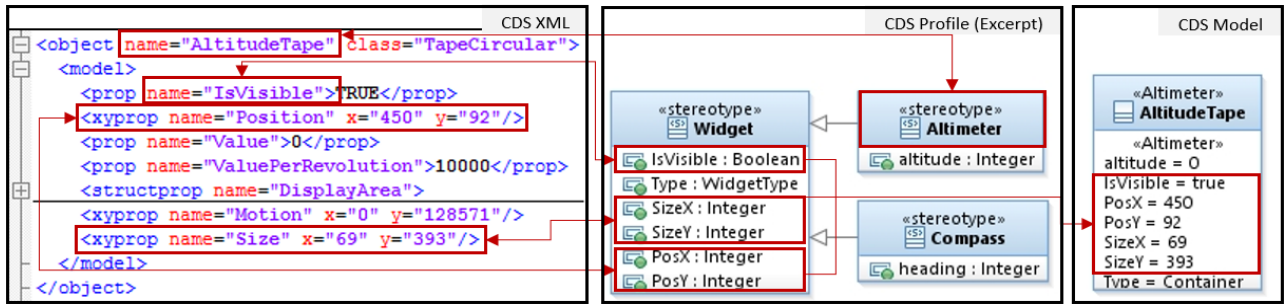

Fig. 3. A mapping between CDS XML and profile and the resulting CDS model

*2) Instance Model Populator:* The main function of this module is to populate the CDS instance model using the data extracted from images. For this purpose, it uses the CDS model generated by the *CDS Model Generator* module. To populate the instance model of CDS, the classes and properties used in the aircraft CDS model are mapped. The data extracted from images for each aircraft CDS class and properties are used to fill the corresponding slots in the instance model.

*3) Constraints Evaluator:* The third sub-module is *Constraints Evaluator*. This part of the tool is mainly used to evaluate the OCL constraints against the data extracted from images. To do so, it uses the CDS model populated by *Instance Model Populator*. The CDS model contains the data obtained from images recorded during the simulation and according to the aircraft flight states. The CDS model is used to prepare an OCL evaluator environment for the evaluation of OCL constraints [10]. Before starting the evaluation process, OCL constraints are loaded from file corresponding to each state of the aircraft flight. The input OCL constraints are evaluated on the CDS instance model. In the case when the data conforms to constraints, the OCL evaluator returns *true* and *false* otherwise. During the evaluation process, the CDS instance model continues to update with the values extracted from images and the results produced by *Constraints Evaluator* are compiled by *Reporting Module*.

### C. Reporting Module

At the end of the evaluation, the *Reporting Module* generates a test report. The report consists of the information regarding passed and failed OCL constraints and the scenarios (i.e, states) in which the faults are encountered. The report also contains the list of OCL constraints that failed for the CDS images. The report generated by this module helps a test engineer to trace the faults in various CDS widgets.

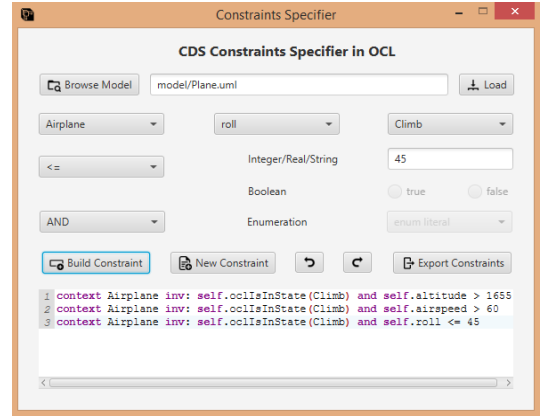

Fig. 4. OCL constraint specifier tool

### D. CDS Constraint Specifier

Typically, avionics engineers are not familiar with OCL terminologies. The lack of OCL knowledge hinders the adaptation of CDST in the industry [11]. Therefore, to assist avionics engineers in writing OCL constraints, we develop a domain-specific tool. Fig. 4 shows the user interface of the tool. The tool allows loading the aircraft model (in UML) containing the flight behavior as a state machine. After loading the model, the aircraft class, properties, and flight states are extracted from the UML model using EMF. All the extracted information required for the OCL constraints is displayed on the user interface (UI). The user can select the class, property, and the flight state for which a constraint is required. Based on the type of selected property, different UI widgets enable or disable to guide the user to enter the correct value. If the type of aircraft property is Integer, all the applicable relational operators are loaded in the combo box and the text box is enabled for Integer value input. For example, Fig. 4 shows that for the *roll* property, relational operator < and the text box with Integer value 45. In the case of Boolean property, the only possible values are *true* and *false*. Thus, two radio buttons for each Boolean value are used. For the enumeration type property, all the enumeration literals are shown in a list. The user can select the desired enumeration literals. The clauses in an OCL constraint are joined with different logical operators (e.g., *and*, *or*, and *implies*). Therefore, the tool provides a list of all logical operators that can be used in an OCL constraint. The complete OCL constraint is built incrementally by adding subparts in the form of constraint clauses. When a constraint is built completely, the user can start creating a new constraint. At the end of writing all constraints, the tool allows exporting the generated OCL constraints in a file. The generated OCL constraints file is provided to the *Comparator* for test evaluation.

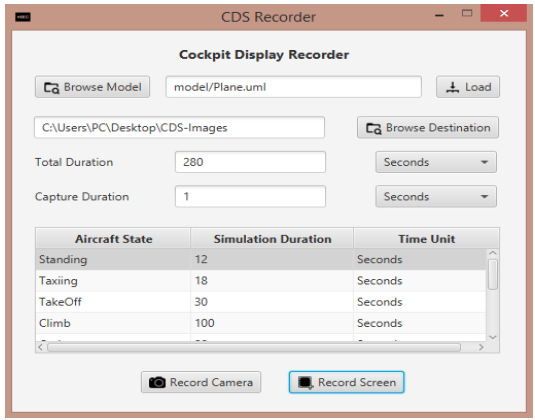

Fig. 5. The user interface of the cockpit display recorder

The constraints used for CDS testing require the information of the aircraft state and the valid range or values of the CDS display properties. For example, the CDS of the aircraft during the *Taxiing* state should display the value of *airspeed* within the range [0, 50]. In this case, the OCL constraints only require *oclIsInState()* OCL operation and primitive properties with relational and logical operations. Therefore, this tool is developed considering the level of complexity required in OCL constraints.

### E. Test Path & Script Generator

To generate tests using the aircraft behavioral model (state machine), *Test Path Generator* implements the N+ test strategy proposed by R. Binder [12]. As an initial requirement, the N+ test strategy needs the flattened state machine. Therefore, this module takes a flattened state machine as input (in UML format). The input UML state machine model is loaded using EMF and is stored in a data structure. The state machine is traversed to achieve round-trip coverage and generate a transition tree. In the case state machine contains cycles, the strategy suggests to allow repetitions one-time only. The generated transition tree is used to generate JSBSim simulator specific test scripts. The simulator scripts are then used to execute tests (using *Test Execution Module*) i.e., to make the aircraft follow the path specified in the transition tree.

### F. Test Execution Module

To execute the test case, it is necessary to interface with a flight simulator. For this purpose, we use JSBSim [13] to simulate the flight dynamics of an aircraft. The test scripts generated by the *Test Path & Script Generator* module for JSBSim simulator are used to execute tests. This module takes all test scripts as input and executes the script to run the simulation.

### G. Cockpit Display Recorder

An important step of the CDS testing approach is to record the display of the cockpit during the simulation according to the specified time interval and aircraft flight states. There are two ways to do that, one is to use an external camera to record the cockpit display during hardware-in-the-loop simulation, and the second way is to record the screenshots of the computer screen on which the (software-in-the-loop) simulation is running. This tool supports cockpit display recording for both cases. Fig. 5 shows the user interface of the tool. The aircraft flight state machine (in UML) is loaded using EMF and the list of flight states is extracted. All the flight states are inserted in the table that allows the user to specify the time duration in which the aircraft will be in the particular state. Besides this, the tool allows the user to provide the total duration of simulation and the time interval after which the image is required to capture. After providing the required information, the user can start the recording either using a camera or taking the screenshots of the computer screen. The images are recorded after the specified interval time and according to aircraft flight states. The recorded images are stored on the hard disk at the specified destination directory. These images are used by the *Data Extractor* module to process each image and extract the relevant information.

## IV. DEMONSTRATION

A major point of interest for the avionics test engineers is the test evaluation time including the time for CDS instance model generation, constraints evaluation, and results reporting. Therefore, the main focus of this demonstration is to analyze the cost of test evaluation in terms of the execution time. In the following, first, we provide the details of the case study used for the demonstration followed by a discussion on demonstration setup, experiment execution, and results.

### A. Case Study

The case study used for the demonstration is developed in collaboration with the CDS development team of our industrial partner using the VAPS XT [6] tool. The case study comprises of the primary flight display (PFD) for an aircraft as shown in Fig. 6. Primary Flight Display (PFD) is the main component of an electronic flight instrument system (EFIS). The Primary Flight Display (PFD) is the primary source of flight information for pilots and displays different types of information like altitude, attitude, airspeed, vertical speed, barometric pressure, and ground speed, etc. Each type of information is shown by a separate graphical widget on the PFD. Thus, PFD is representative of a real CDS because it composes the information displayed on individual widgets such as an Altimeter to display altitude and an Airspeed Indicator to show the airspeed.

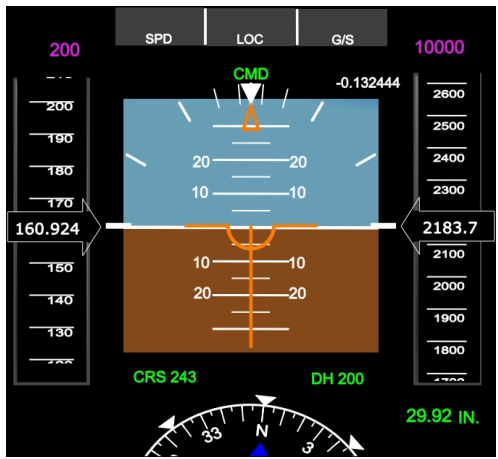

Fig. 6. A screen grab of a Primary Flight Display (PFD)

TABLE I
IMAGES RECORDED DURING SIMULATION, AVERAGE EVALUATION TIME, IMAGE PROCESSING TIME, FAILED OCL EVALUATIONS, AND UNIQUE OCL CONSTRAINTS FAILED

| State | Images | Avg. Eval. Time (m) | Img. Proc. Time (m) | Failed OCL Eval. | Unique OCL Failed |
|---|---|---|---|---|---|
| **Standing** | 300 | 2.62649 | 2.92 | 0 | 0 |
| **Taxiing** | 1248 | 9.26912 | 13.65 | 824 | 2 |
| **TakeOff** | 460 | 1.50735 | 4.09 | 83 | 2 |
| **Climb** | 4660 | 44.91825 | 44.44 | 428 | 1 |
| **Cruise** | 4258 | 14.47 | 43.04 | 219 | 1 |
| **Descent** | 3462 | 36.30 | 35.2338 | 320 | 2 |
| **StraightAnd-Level** | 1410 | 1.188 | 13.8562 | 515 | 2 |
| **Approach** | 300 | 2.66463 | 2.99 | 195 | 1 |
| **Landing** | 402 | 3.57061 | 4.02 | 261 | 1 |
| **Total** | **16500** | **116.5196** | **164.2683** | **2854** | **12** |

A simulation of PFD of an aircraft flying at 2183 feet above sea level (ASL) is shown in Fig. 6. On the left side of the PFD, there is an airspeed tape that shows the airspeed of the aircraft. In Fig. 6 the airspeed is ≈160 knots. On the right-hand side of PFD, there is an altitude tape showing the altitude of the aircraft, i.e., ≈2183 feet ASL. The center of the PFD contains the attitude indicator that shows the pitch and roll of the aircraft. Barometric pressure is shown in green color below the altitude tape on the bottom right corner.

### B. Demonstration Setup

All the structural details of PFD (i.e, the location and relative scales of various widgets) are present in the XML file generated by the VAPS XT tool. The PFD XML file is used to create the CDS model as an instance model of the CDS profile. For the behavioral model of an aircraft, we use the reference state machine presented in the previous work [5]. The aircraft flight state machine is used to generate test paths using *Test Path Generator* tool.

To execute the test cases (paths), it is necessary to interface with a flight simulator. We use JSBSim [13] to simulate the data for various widgets obtained from the flight dynamics model of Cessna 172 Skyhawk aircraft. During the simulation, at each aircraft state during the flight, images are recorded after one second and stored with respect to the state. The total number of simulation scripts executed for this demonstration is 20. The test paths used for simulation contain all aircraft states necessary for complete flight. The simulation data is available at the Github repository[3]. The expected properties of the widgets for the aircraft states are modeled as OCL constraints. There are 24 distinct OCL constraints on the various PFD widgets that are identified during different sessions with our industry partner.

[3]https://github.com/hassansartaj/cdst-toolkit/datasets

### C. Experiment Execution

The purpose of the demonstration is to analyze the execution time of test evaluation. To calculate execution time, we perform 10 individual experiment trials of test evaluation using the same settings (i.e., same simulation data and OCL constraints). The aim of multiple experimental trials is to analyze the average execution time the CDST takes during test evaluation. The execution time is analyzed for each aircraft flight state individually. We use one machine to execute the experiment. The specifications of the machine are 3.2 GHz core i7 processor, 32GB RAM, 1 TB hard drive, and Windows 10 operating system.

### D. Results and Discussion

Table I shows the results of 200 test case executions in which 20 different test scenarios are executed for 10 times each. The table shows the number of images recorded during simulation, average evaluation time for each test executions, image processing time, failed OCL constraints evaluations, and the number of unique constraints failed. The data is shown for each aircraft flight state.

The total test execution time for 200 test executions is 2750 minutes. The average execution time for each test case is ≈13.75 minutes The total test evaluation time for 200 test executions is approximately 19 hours. The average evaluation time for each test case is ≈1.9 hours. Our evaluation shows that 12 OCL constraints were violated for every execution of 20 test scenarios. Manual analysis shows that these constraints are mapped to 3 distinct faults in the PFD. The three faults are attributed to the three parts of PFD, i.e., airspeed tape, altimeter tape, and attitude indicator. The corresponding OCL constraints are failed for airspeed, altitude, and roll values.

The evaluation results are based on execution on a single machine, however, in practice, the image processing can easily be done in parallel. This will reduce the offline overhead of test evaluation significantly. According to our experience with avionics test engineers, for the evaluation of the same amount and type of test scenarios, the manual testing of CDS usually takes more than one week. Therefore, the time that CDST takes for the complete test evaluation is much less as compared to manual testing done by avionics test engineers.

## V. LIMITATIONS

An important step in our approach is to use image processing to extract relevant information (e.g., text) from various CDS widgets. The prediction accuracy of the OCR engine such as Tesseract [9] poses a limitation to our toolkit. The accuracy of Tesseract OCR is not always 100% [14], [15]. To handle this limitation and to enhance the accuracy, we used region-based segmentation and image preprocessing techniques such as noise removal, canny edge detection, and contours finding.

Currently, the CDST toolkit supports the CDS XML file generated by the VAPS XT tool. The VAPS XT tool is widely used in the avionics domain for designing the CDS. Therefore, the current version works for the CDS designed using the VAPS XT tool. However, an interface is available to add support for the XML file generated from the SCADE tool. Moreover, the CDST toolkit supports simulator scripts generation for the JSBSim simulator which is widely used for simulation and testing [16], [17]. In the future, we plan to add support for the generation of scripts for a variety of aircraft flight simulators.

## VI. RELATED WORK

A number of GUI testing tools are available for desktop and web applications, including GUITAR [18], Sikuli [19], Sikuli Test [20], JAutomate [21], Android Ripper [22], Amola [23], and Orbit [24]. However, CDS testing requires interaction with propriety Multi-functional displays, ARINC 661 [25] compliance, and during testing the aircraft behavior needs to be simulated.

## VII. CONCLUSION

An important step in testing the user application is to test whether the required information is being displayed correctly on the Cockpit Display Systems (CDS) of an aircraft. The current industrial practice is to test this manually, which is very labor extensive and error-prone. In this paper, we present an initial version of the CDST toolkit to automate the testing of Cockpit Display Systems (CDS). The toolkit consists of various components to assist the avionics engineers in the process of testing CDS. We demonstrate the tool on an industrial case study. The results show that the average evaluation time for a test case is ≈1.9 hours. The results also show that the tool is able to identify 3 distinct faults in the case study. In the future, we plan to perform a pilot study to evaluate the usability of the *OCL Constraint Generator* tool using avionics engineers as subjects.